\documentclass[pra,twocolumn,amsmath, superscriptaddress,notitlepage,showpacs,10pt]{revtex4-1}
\usepackage{amsmath,amsfonts,amssymb,amsthm,amstext,amscd,amsthm,dsfont,bbm,hyperref,natbib}
\usepackage{physics}
\usepackage{color}
\usepackage{soul,xcolor}
\usepackage{gensymb}
\hypersetup{
    colorlinks = true,
    citecolor=blue,
    linkcolor = red,
    anchorcolor = red,
    citecolor = blue,
    filecolor = red,
    pagecolor = red,
    urlcolor = blue,
    }
\usepackage[normalem]{ulem}
\usepackage{graphicx}
\usepackage{bbold}
\usepackage{braket}
\usepackage{placeins}
\newtheorem{theorem}{Theorem}

\newtheorem{Definition}{Definition}

\begin{document}

\title{Compressed Sensing Tomography for qudits in Hilbert spaces of non-power-of-two dimensions}
	\author{Revanth Badveli}
	\email{badveli.revanth@gmail.com}
	\affiliation{
		Computer Science and Information Systems, BITS Pilani-Goa Campus, Goa 403 726, India}
	\affiliation{Quantum Research Group, School  of Chemistry and Physics,
		University of KwaZulu-Natal, Durban 4001, South Africa}
	\author{Vinayak Jagadish}
	\email{jagadishv@ukzn.ac.za}
	\affiliation{Quantum Research Group, School  of Chemistry and Physics,
		University of KwaZulu-Natal, Durban 4001, South Africa}\affiliation{ National
		Institute  for Theoretical  Physics  (NITheP), KwaZulu-Natal,  South
		Africa}
			\author{R. Srikanth}
	\affiliation{Poornaprajna Institute of Scientific Research,
		Bangalore- 560 080, India}
	\author{Francesco Petruccione}
	\affiliation{Quantum Research Group, School  of Chemistry and Physics,
		University of KwaZulu-Natal, Durban 4001, South Africa}\affiliation{ National
		Institute  for Theoretical  Physics  (NITheP), KwaZulu-Natal,  South
		Africa}
\date{} 

\begin{abstract} 
		The techniques of low-rank matrix recovery were adapted for Quantum State Tomography (QST) previously by Gross {\it et al.} [Phys. Rev. Lett. 105, 150401 (2010)], where they consider the tomography of $n$ spin-$1/2$ systems. For the density matrix of dimension $d = 2^n$ and rank $r$ with $r \ll 2^n$, it was shown that randomly chosen Pauli measurements of the order $O[dr \log(d)^2]$ are enough to fully reconstruct the density matrix by running a specific convex optimization algorithm.  The result utilized the low operator-norm of the Pauli operator basis, which makes it ``incoherent" to low-rank matrices.  For quantum systems of dimension $d$ not a power of two, Pauli measurements are not available, and one may consider using SU($d$) measurements. Here, we point out that the SU($d$) operators, owing to their high operator norm, do not provide a significant savings in the number of measurement settings required for successful recovery of all rank-$r$ states.  We propose an alternative strategy, in which the quantum information is swapped into the subspace of a power-two system using only $\textrm{poly}[\log(d)^2]$ gates at most, with QST being implemented subsequently by performing $O[dr \log(d)^2]$ Pauli measurements. We show that, despite the increased dimensionality, this method is more efficient than the one using SU($d$) measurements.
\end{abstract}
\maketitle  
\section{Introduction}
Quantum state (process) tomography \cite{paris_quantum_2004} is the procedure of experimentally characterizing an unknown quantum state (process). It is an increasingly important task in quantum information processing \cite{nielsen_quantum_2000}. To characterize an unknown $d$ dimensional quantum state, one would need to estimate the expectation of values of a set of $d^2$ observables, which span the space of $d \times d$ Hermitian matrices. To characterize a quantum process acting on a $d$-dimensional quantum system, one would need to input $d^2$ linearly independent quantum states to the process and do a state tomography on all $d^2$ outputs. This is due to the fact that the output of a quantum process for any unknown arbitrary input state can be determined by it's action on a set of linearly independent quantum states whose density matrices span the space of $d\times d$ matrices.
The main problem associated with any quantum tomography task  is that the dimension of the system grows exponentially with it's size, making the whole task resource intensive.  

One can hope to reduce the measurement settings by restricting the classes of states (processes) subject to characterization. For example, if a  process matrix \cite{jagadish_invitation_2018,nielsen_quantum_2000} of an unknown quantum process acting on a $d$-dimensional quantum system is known to be $s$-sparse in certain known basis then it is shown in Ref. \cite{shabani_efficient_2011} that compressed sensing (CS) techniques \cite{candes_stable_2006,donoho_optimally_2003,baraniuk_simple_2008} can be adapted to characterize the process matrix using $O[s \log(d)]$ measurement settings. This method was experimentally performed for a two qubit gate in Ref. \cite{shabani_efficient_2011} and for superconducting quantum gates in Ref. \cite{rodionov_compressed_2014}. Similar techniques are used in Refs. \cite{shabani_estimation_2011, rudinger_compressed_2015} to characterize $s$-sparse Hamiltonian (in known basis) of $d$-dimensional systems using only $O[s \log(d)]$ measurement settings. 

 The matrix generalization of CS techniques, known as matrix completion \cite{candes_exact_2009,recht2011simpler,recht_guaranteed_2010}, are adapted to quantum state tomography (QST) by Gross {\it et al.} \cite{gross_quantum_2010} where they consider tomography of $n$ spin-$1/2$ systems, whose density matrix $\rho$ is of dimension $d = 2^n$ and rank-$r$. It was shown that $\abs{\Omega} = cdr\log(d)^2$ randomly chosen Pauli measurements are enough to recover $\rho$ with exponentially low failure probability in $c$ by running a certain convex optimization algorithm. Numerical performance and robustness of these methods to noise are discussed in Ref. \cite{flammia_quantum_2012}. The experimental implementation of these methods are presented in Refs. \cite{riofrio_experimental_2017,steffens_experimentally_2017,liu_experimental_2012}. Similar results were obtained in Ref. \cite{liu_universal_2011} by making use of the restricted isometry property (RIP). CS-QST protocol using continuous measurements on unknown low-rank quantum states, which is being manipulated by controlled external fields, is presented in \cite{smith_quantum_2013}. In \cite{kyrillidis_provable_2018}, a non convex algorithm is proposed for CS QST setting to improve the running time.  In general, tomography of unknown quantum states restricted by prior information is studied in \cite{heinosaari_quantum_2013}.
 
The main results of Ref. \cite{gross_quantum_2010} were generalized to any given matrix basis in Ref. \cite{gross_recovering_2011} where it is shown that $O[dr\nu \log(d)^2]$ expectation values with respect to the given operator basis are sufficient to recover an unknown rank-$r$, $d$ dimensional quantum states. The number $\nu$ is the "coherence" of the density matrix with respect to the given matrix basis. Note that the coherence $\nu$, which is defined later in the article, is different from the quantum coherence \cite{baumgratz_quantifying_2014}. The result of Ref. \cite{gross_quantum_2010} follows from Ref. \cite{gross_recovering_2011} due to the fact that all the low-rank matrices have coherence $\nu = 1$ with respect to Pauli operator basis. However, for quantum systems of dimension $d$ not a power of two, one cannot perform Pauli measurements.

A natural option would be to measure SU($d$) generators  \cite{greiner_quantum_1994}, which from here on will be referred to as SU($d$) measurements. The set of SU($d$) generators are the natural extension of Pauli matrices to $\mathbbm{C}^{d \times d}$. This set consists of $d^2-1$ traceless, orthonormal, Hermitian operators and the identity operator.

 We find that SU($d$) measurements do not guarantee "universal recovery" due to it's high `coherence' \cite{gross_recovering_2011}. We propose an alternative strategy, in which the quantum information is transferred from the system to a power-two ancilla using a unitary operation $W$, which can be efficiently implemented using $\textrm{poly} [\log(d), 1/\epsilon]$ gates with accuracy $\epsilon$. CS-QST is then performed using $\abs{\Omega} = c'dr\log(d)^2$ randomly chosen Pauli measurements on the ancilla to reconstruct the density matrix of the unknown quantum state. We further compare the performance of this method with the one where SU($d$) measurements are used. Certain quantum communication tasks have increased security against the attacks when qutrits and higher dimensional states are used \cite{bechmann-pasquinucci_quantum_2000,macchiavello_security_2003,durt_security_2004} and reconstructing such states can be of particular interest which validates the necessity for considering systems of dimensions not a power of two.

The outline of the paper is as follows. The required definitions and notations are introduced in Sec. \ref{preliminaries}. In Sec. \ref{Sud} we discuss the problems arising from the usage of SU($d$) measurements for reconstruction. An alternate method is discussed in Sec. \ref{AltStrat}. In Section \ref{Swap} we discuss the gate complexity for a unitary operation introduced in our methods. Finally, we conclude in Sec. \ref{conclusion}.
\section{Preliminaries}
\label{preliminaries}
We use three matrix norms in this article, namely the nuclear norm, the Frobenius norm and the operator norm. Consider a $d\times d$ matrix $X$. 
\begin{Definition}[Nuclear norm]
	\label{def}
The nuclear norm of $X$ is given as $\norm{X}_1 = \sum_{i}^{d} \sigma_i (X)$, where $\{ \sigma_i (X) \}$ are the singular values of $X$.
\end{Definition}
\begin{Definition}[Frobenius norm]
The Frobenius norm of $X$ is defined as
$\norm{X}_2 = \mathrm{Tr}(X^\dagger X) = \sqrt{\sum_{i}^{d} \sigma_i (X)^2}$.
\end{Definition}
\begin{Definition}[Operator norm]
 The operator norm is evaluated as $\norm{X} = \max_i [\sigma_i (X)]$. 
\end{Definition}

Following Ref. \cite{gross_recovering_2011}, we refer to an orthonormal basis  $\{ w_a\}_{a=1}^{d^2} $ with respect to the inner product $\braket{X, Y} = \mathrm{Tr}(X^\dagger Y)$ in the space of $d\times d$ matrices, where each element is Hermitian ($w_a = w_a^\dagger$), as the operator basis. Any $\rho$ ($d\times d$) can be expanded as
\begin{equation}
\rho = \sum_{a=1}^{d^2} \braket{w_a,\rho} w_a.
\end{equation}
Each expansion coefficient $\braket{w_a,\rho}$ can be interpreted as the expected value of the observable $w_a$ on $\rho$.

\begin{Definition}[Coherence]\label{def:coherence}The coherence $\nu$ of a $d \times  d$ matrix $\rho$ with respect to an operator basis $\{ w_a\}_{a=1}^{d^2}$ is given by $\min( \nu_1, \nu_2)$ if 
\begin{equation}
\max_a \norm{w_a}^2 \leq \nu_1\left(\frac{1}{d}\right)
\end{equation} and
\begin{equation}
\max_a \norm{P_U w_a + w_a P_U - P_U w_a P_U}_2^2 \leq 2\nu_2 \left(\frac{r}{d}\right)
\end{equation}
hold. $P_U$ is the projection operator onto the column (or row) space of $\rho$. 
\end{Definition}
Note that $\nu_1$ is independent of the density matrix $\rho$ unlike $\nu_2$.

\begin{theorem}
		\label{Thm:mainthm2} See Reference \cite{gross_quantum_2010}.
			Let $\rho$ ($d \times d$) be an arbitrary state of rank $r$. Let $\Omega \subset \{w_a\}_{a= 1}^{d^2}$ be a randomly chosen set. Each operator $w_a$ is a k-fold tensor product of the Pauli basis operators $\{\sigma_i\}_{i=0}^3$ for matrices on $(\mathbb{C}^2)^{\otimes k}$, where $d^2 = 2^k$. If the number of Pauli expectation values $m = \abs{\Omega} = c d r \log(d)^2$ then the solution $\sigma^*$ to the following optimization program, 
			\begin{eqnarray}
			\label{eq: mainprog2}
			&\min \;\; \norm{\sigma}_1 \nonumber \\
			&\mathrm{subject\thinspace to} \;\; \mathrm{Tr}(w_a\sigma) =  \mathrm{Tr}(w_a\rho) \; \forall w_a \in \Omega,
			\end{eqnarray}
			 is unique and equal to $\rho$ with failure probability exponentially small in $c$. 
	\end{theorem}

\begin{theorem}
	\label{Thm:mainthm} See Reference \cite{gross_recovering_2011}.
	Let $\rho$ $(d \times d)$ be a rank-$r$ matrix with coherence $\nu$ with respect to the operator basis $\{w_a\}_{a= 1}^{d^2}$. Let $\Omega \subset \{w_a\}_{a= 1}^{d^2}$ be a randomly chosen set. The solution $\sigma^*$ to the following optimization program,
	\begin{eqnarray}
	\label{eq: mainprog}
	&\min \;\; \norm{\sigma}_1 \nonumber \\
	&\mathrm{subject\thinspace to} \;\; \mathrm{Tr}(w_a\sigma) =  \mathrm{Tr}(w_a\rho) \;\; \forall w_a \in \Omega,
	\end{eqnarray}
	 is unique and equal to $\rho$ with probability of failure smaller than
	$e^{-\beta}$ provided that 
	\begin{equation*}
	\abs{\Omega} \geq O[dr\nu (\beta+1)\log(d)^2].
	\end{equation*}
\end{theorem}

\section{SU($d$) Operator basis} 
\label{Sud}
Consider the tomography of $n$ spin-$1/2$ systems, whose density matrix is of dimension $d = 2^n$ and rank-$r$. Gross {\it et al.} \cite{gross_quantum_2010} show that $cdr\log(d)^2$ randomly chosen Pauli measurements are sufficient to reconstruct the density matrix from program (\ref{eq: mainprog}) with exponentially low failure probability in $c$. The operator norm of any normalized Pauli operator is $\sqrt{1/d}$, and hence, $\nu_1$ = 1. For any given density matrix, the number $\nu_2$ is also equal to one with respect Pauli operator basis due to,
\begin{eqnarray}
\max_a \norm{P_U w_a + w_a P_U - P_U w_a P_U}_2^2 &\leq& \sup_{\sigma \in \mathcal{T}, \norm{\sigma}_2 = 1} \braket{w_a,\sigma} \nonumber \\ 
&\leq& \norm{w_a}^2\norm{\sigma}_2^2 \nonumber \\ 
&\leq& \norm{w_a}^2 2r \norm{\sigma}_2^2 \nonumber \\
&\leq& \frac{2r}{d},
\end{eqnarray}
where $P_U$ is the projector onto the column space of the density matrix and $\mathcal{T}$ is the set of matrices ($Y$) which satisfy the condition $ (\mathbbm{1} - P_U)Y(\mathbbm{1} - P_U) = 0$ \cite{gross_recovering_2011}.
With respect to the Pauli operator basis, the coherence of any density matrix is $\nu = \nu_1 =\nu_2 = 1$. Hence the result in Ref. \cite{gross_quantum_2010} follows straight forwardly from Theorem \ref{Thm:mainthm}.

Let us now consider the task of reconstructing rank-$r$ quantum states of dimension ($d$) not a power of two using the techniques given in Ref. \cite{gross_quantum_2010,gross_recovering_2011}.

 Since the Pauli operator can only be defined in $\mathbbm{C}^{2^k \times 2^k}$ as a $k$-fold tensor product of $SU(2)$ operators, a natural candidate would be to use the SU($d$) operator basis \cite{greiner_quantum_1994}. The operator norm of SU($d$) basis elements is greater than or equal to $1/2$, and hence, $\nu_1 > d/2$. In this case, one can obtain non-trivial bounds on the number of SU($d$) measurement settings from Theorem \ref{Thm:mainthm} only if $\nu_2$ is small. From the definition of $\nu_2$,

\begin{align}
 &\max_a \norm{P_U w_a + w_a P_U - P_U w_a P_U}_2^2\nonumber\\
 &=\max_a\; 2 \braket{P_U w_a,P_U w_a}  -  \braket{P_Uw_a P_U,P_U w_a P_U}\nonumber\\
 &\leq  \max_a\; 2 \braket{P_U w_a,P_U w_a} \nonumber\\
 &=  \max_a\; 2\thinspace\mbox{Tr}(P_U w_a^2). 
 \label{ineq:coherence2}
 \end{align}
Observe that $ w_a^2$ is a diagonal matrix for all $w_a \in$ SU($d$).
If we restrict our attention to pure quantum states [i.e. rank$(\rho) = 1$] then the inequality (\ref{ineq:coherence2}) can be reduced to $\max_{i,j, i \neq j}\;\rho_{ii} + \rho_{jj}$. So the bounds obtained from Theorem \ref{Thm:mainthm} are non trivial when $\max_{i,j, i \neq j}\;\rho_{ii} + \rho_{jj}$ is small, much like the coherence condition in Ref. \cite{candes_exact_2009}. For example, consider the task of performing CS-QST using $SU(7)$ operator basis on following quantum states,
\begin{eqnarray}
	\rho_1 &=& \ketbra{0}\\ \nonumber
	\rho_2 &=& \frac{1}{7} \sum_{i,j = 0}^{6} \ketbra{i}{j},
\end{eqnarray}
where $\{\ket{i}\}_{i =0 }^{6}$ for the standard basis for $\mathbbm{C}^7$. With respect to the SU($d$) basis, the state $\rho_1$ has the maximum coherence, whereas $\rho_2$ has the minimum coherence. A numerical simulation reveals that one can exactly reconstruct $\rho_1$ only  $95\%$ times from $46$ $SU(7)$ measurements chosen uniformly at random, whereas $\rho_2$ can be exactly reconstructed the same number of times using only $28$ $SU(7)$ measurement settings chosen uniformly at random. This shows that one can gain advantage by performing CS-QST using SU($d$) measurements only when the number $ \max_a\; \thinspace\mbox{Tr}(P_U w_a^2)$ is small, which may not be possible to know beforehand. This issue of operator norm with respect to the SU($d$) generators, therefore indicates that they are not the best candidates as measurement operators. We, therefore, propose an alternate method in the next section to overcome this problem. 

\section{Alternate approach}
\label{AltStrat}
From Theorem \ref{Thm:mainthm}, it is clear that if there exists an operator basis $\{w_a\}_{a =1}^{d^2}$ with small $\nu_1$ in the space of $d \times d $ Hermitian matrices where $d$ is not a power of two, one can recover any quantum state from only $O[d r \log(d)^2]$ measurement settings. Instead of searching for such an operator basis, we propose a method where we transfer the quantum information from the system to the ancilla efficiently. We then perform CS-QST on the ancilla using Pauli measurements. This strategy also gives us the advantage of employing Pauli measurements which are more easily implementable than SU($d$) measurements. 

Let the system $\rho_S$ be a rank-$r$ density matrix acting on $\mathbbm{C}^{d_1}$, where $d_1$ is not a power of two, and the ancilla $\rho_A$ is acting on $\mathbbm{C}^{d_2}$. The dimension of the ancilla $d_2$ is set to a power of two greater than $d_1$. This is because we would like to perform Pauli measurements on the ancilla $\rho_A$ at a later stage. The system is first coupled unitarily to the ancilla by a swap operator  $W$,
\begin{equation}
\rho_{SA} = W \rho_S \otimes \rho_A W^\dagger.
\end{equation}
For our purposes we define $W$ as the following,
\begin{equation}
W =	\sum_{i,j}^{d_1} \ketbra{i_S}{j_S}\otimes \ketbra{j_A}{i_A} + \sum_{i}^{d_2 -d_1} \mathbbm{1} \otimes \ketbra{i_A},
\end{equation}
where $\{ i_S\} $ and $\{ i_A\} $ form the orthonormal basis in $\mathbbm{C}^{d_1}$ and  $\mathbbm{C}^{d_2}$ respectively. It swaps the $d_1 $-dimensional space of the system with $d_1$-dimensional subspace of the ancilla which is spanned by $\{\ket{i_A}\}_{i = 0}^{d_1}$. Let the initial state of the system $\rho_S$ be $\sum_{i, j}^{d_1} \rho_{ij} \ketbra{i_S}{j_S}$.  One can choose the initial state of the ancilla from the $d_1$ dimensional subspace spanned by $\{\ket{i_A}\}_{i = 0}^{d_1}$.
For brevity of analysis, we set the initial state to,
\begin{equation}
\label{Eq:minwithanc}
\rho_A = \ketbra{0_A} =  \mqty(1 & 0 & \dots & 0\\
0 & 0 & \dots & 0\\
\vdots &\vdots&\ddots&\vdots\\
0 & 0 & \dots & 0)_{d_2 \times d_2}.
\end{equation}
The combined state of system + ancilla after the action of unitary $W$ is,
\begin{align}
&\rho_{SA} = W \rho_S \otimes \rho_A W^\dagger \nonumber \\ 
&= \left(\sum_{i', j'}^{d_1} \ketbra{i'_S}{j'_S}\otimes \ketbra{j'_A}{i'_A} + \sum_{i'}^{d_2 -d_1} \mathbbm{1} \otimes \ketbra{i'_A} \right) \nonumber\\
&\times \left(\sum_{i,  j}^{d_1} \rho_{ij} \ketbra{i_S}{j_S} \otimes \ketbra{0_A} \right) W^\dagger  \nonumber\\
&= \left( \sum_{j', i, j }^{d_1} \rho_{ij} \ket{0_S} \braket{j'_S|i_S}\bra{j_S} \otimes \ketbra{j'_A}{0_A}  \right)W^\dagger \nonumber \\
&= \left( \sum_{i, j }^{d_1} \rho_{ij} \ket{0_S}\bra{j_S} \otimes \ketbra{i_A}{0_A}  \right)  \nonumber\\
& \times\left( \sum_{i', j'}^{d_1} \ketbra{j'_S}{i'_S}\otimes \ketbra{i'_A}{j'_A} + \sum_{i'}^{d_2 -d_1} \mathbbm{1} \otimes \ketbra{i'_A}  \right)\nonumber \\
&= \sum_{j', i, j }^{d_1} \rho_{ij} \ket{0_S} \braket{j_S| j'_S}\bra{0_S}  \otimes \ketbra{i_A}{j'_A} \nonumber \\
&= \ketbra{0_S} \otimes\left(\sum_{ i, j }^{d_1} \rho_{ij} \ketbra{i_A}{j_A} \right).
\end{align}

One can see that the new state of the ancilla $\rho'_A = \sum_{ i, j }^{d_1} \rho_{ij} \ketbra{i_A}{j_A} $ has $\rho_S$ on the top left $d_1 \times d_1$ block and zeros elsewhere. This implies that the $\mbox{rank} (\rho'_A) = \mbox{rank} (\rho_S)$, and one can recover $\rho'_A $ using CS-QST to get $\rho_{S}$. We use the following program to reconstruct $\rho'_A$,
\begin{eqnarray}
\label{eq: altprog}
&\min \;\; \norm{\sigma}_1 \nonumber \\
&\mathrm{subject \thinspace to} \;\;\; \mbox{Tr}(w_a\sigma) =  \mbox{Tr}(w_a\rho'_A) \; \forall w_a \in \Omega,
\end{eqnarray}
where $\Omega$ is the set of randomly chosen Pauli operators. From Theorem \ref{Thm:mainthm2}, it directly follows that $\abs{\Omega} = c d_2 r \log(d_2)^2$ Pauli measurements are enough for the output of the program (\ref{eq: altprog}) to be unique and equal to $\rho'_A$ with failure probability exponentially low in $c$.
To reduce the number of measurement settings, we set $d_2$ as the smallest power of two greater than or equal to $d_1$. The number of measurement settings $ c d_2 r \log(d_2)^2$  can then be upper bounded by  $ c' d_1 r \log(d_1)^2$ as the $d_2$ is always less than $2 d_1$.

\begin{figure}
	\begin{center}
		\includegraphics[scale = 0.41]{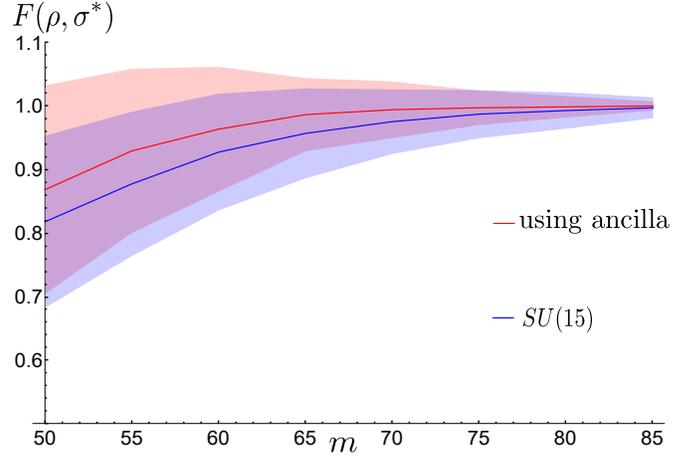}\hspace{5pt}
		\caption{The fidelity $F(\rho,\sigma^*)$ between the estimated ($\sigma^*$) and the true states ($\rho$) against the number of measurement settings ($m$) for SU(15) basis measurements (orange) and Pauli measurements on the ancilla (blue) is shown. Fidelity is calculated over 2000 randomly generated $15\times 15$ rank-1 density matrices.} \label{fig1}
	\end{center}
\end{figure}

\begin{figure}
	\begin{center}
		\includegraphics[scale = 0.41]{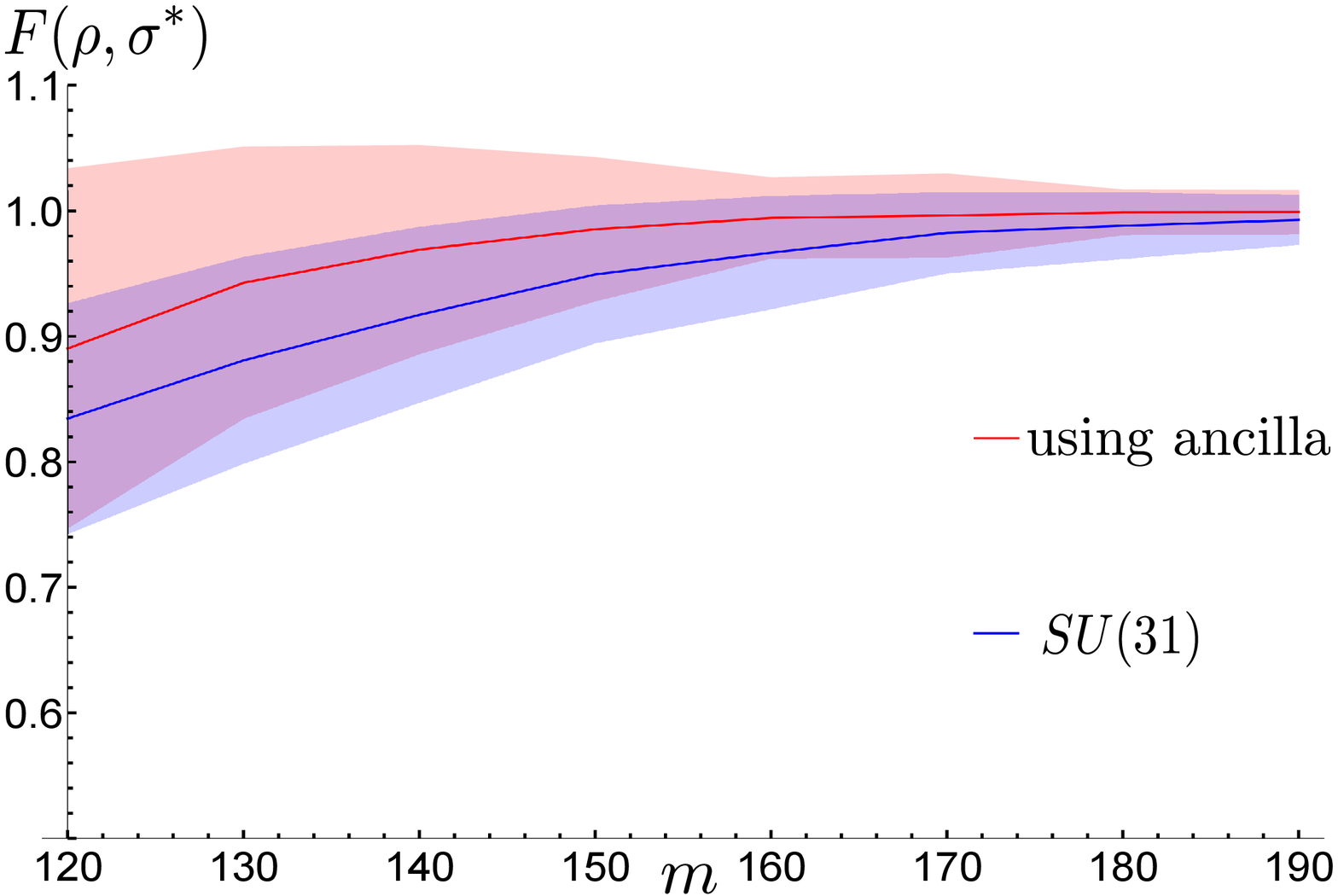}\hspace{5pt}
		\caption{The fidelity $F(\rho,\sigma^*)$ between the estimated ($\sigma^*$) and the true states ($\rho$) against the number of measurement settings  ($m$) for SU(31) basis measurements (orange) and Pauli measurements on the ancilla (blue) is shown. Fidelity is calculated over 1000 randomly generated $31\times 31$ rank-1 density matrices.} \label{fig2}
	\end{center}
\end{figure}

We performed numerical simulations to compare the performance of the alternate approach with the one using SU($d$) measurements. The simulations were performed in MATLAB using a freely available package \cite{cvx}. The simulations although noiseless, are sufficient to bring out the main ideas that we present. In Fig. \ref{fig1}, we compare the Fidelity, which is defined as $F(\rho,\sigma^*) = \Tr(\sqrt{\sqrt{\rho} \sigma^* \sqrt{\rho}})^2$,  between the estimated ($\sigma^*$) and true states ($\rho$) against the number of measurement settings for SU(15) basis measurements (blue) and Ancilla aided approach (orange). Fidelity is calculated over 2000 randomly and uniformly generated $15\times15$ rank-1 density matrices. One can see that the performance using the ancilla aided approach is better for all the considered measurement settings. In Fig. \ref{fig2}, we compare the fidelity between the estimated and the true states against the number of measurement settings for SU(31) basis measurements (blue) and the alternate approach (orange). Fidelity is calculated over 1000 randomly and uniformly generated $31\times31$ rank-1 density matrices. As we increase the dimension of the density matrices, we see that the difference in the performance becomes more apparent because the number of measurement settings for the alternate approach scale better than the one using SU($d$) measurements. Note that in Figs. \ref{fig1} and \ref{fig2}, the shaded regions cover the region between the sum and difference of the mean and standard deviation (mean $ \pm $ standard deviation) of $F(\rho,\sigma^*)$ for a given measurement setting.

\section{Gate Complexity of $W$}
\label{Swap}
The sparsity of the unitary operator $W$ makes it efficiently implementable using only single qubit gates. It is shown, in Refs. \cite{jordan_efficient_2009,childs2004quantum}, that one can implement any unitary $U$ by evolving the system under the Hamiltonian $\mqty(0 & U \\ U^\dagger & 0)$.  Furthermore, according to Ref. \cite{aharonov_adiabatic_2003}, if a $N \times N$ Hamiltonian $H$ has at most $d$ non-zero entries in every row, one can implement it with an error $\epsilon$ using $\textrm{poly}\; [\log(N), d, \norm{Ht}, 1/\epsilon]$ gates.
Following \cite{jordan_efficient_2009,childs2004quantum}, let 
\begin{equation}
	H = \mqty( 0 & W \\ 
			 W^\dagger & 0).
\end{equation} 
One can see that $H$ is one-row-sparse as $W$. Using Taylor series expansion one can write $e^{-i H t}$ as
\begin{equation}
	e^{-\imath H t} = \cos(t) \mathbbm{1} - \imath \sin(t) H.
\end{equation}
By choosing $t$ appropriately, one can get 
\begin{equation}
	e^{-\imath H t} = - \imath H = - \imath \mqty( 0 & W \\ 
	W^\dagger & 0) = - \imath\sigma_x \otimes W. 
\end{equation}
The Hamiltonian $H$ generates the following evolution,
\begin{equation}
	e^{-\imath H t} (\rho_f \otimes \rho_S \otimes \rho_A) e^{+ \imath H t} = -\imath \sigma_x \rho_f\sigma_x \otimes W \rho_S \otimes \rho_A W^\dagger, 
\end{equation}
where $\rho_f$ is a qubit in the first register which can be ignored after the computation. To implement the $d_1 d_2 \times d_1 d_2$ unitary matrix $W$ with an error less than $\epsilon$, one would need  $\textrm{poly}[\log(d_1 d_2), 1/\epsilon]$ gates. One can upper bound the number of gates required by  $\textrm{poly}[\log(d_1), 1/\epsilon]$ using $d_2 \leq 2 d_1$.
\section{Discussion and conclusions}
\label{conclusion}

In this article, we consider the problem of performing CS-QST on quantum systems of dimension not a power of two. For power-two systems, it is shown in Ref. \cite{gross_quantum_2010} that one needs $O[d r \log(d)^2]$ Pauli expectation values where $r$ and $d$ are rank and dimension of the system's density matrix respectively. The result makes use of the low operator-norm of the Pauli basis, which is applicable only on Hilbert spaces whose dimension is a power of two. To achieve the same asymptotic bounds for the considered problem,  we proposed an alternate approach, which uses Pauli measurements and requires relatively less additional cost when compared to the cost of performing CS-QST. In this approach, we transfer the quantum information in the system to an ancilla of power-two dimension using a general unitary operation $W$, which can be implemented with accuracy $\epsilon$ using at most $\textrm{poly}\; [\log(d_1), 1/\epsilon]$ gates. We showed that $ c^\prime d_1 r \log(d_1)^2$ random Pauli measurements on the ancilla are enough to exactly recover the density matrix of quantum states using the convex optimization algorithm (\ref{eq: mainprog}). The performance of the proposed method is shown to be better than the one where SU($d$) measurements are used.  How this performance can be improved by applying efficiently implementable pseudo-unitary on the ancilla ahead of Pauli measurements is a part of future research. The methods introduced in the article can be extended to quantum process tomography by performing CS-QST on the Choi-Jamio{\l}kowski state \cite{jamiolkowski_effective_1974,choi_completely_1975} $\rho_{\mathcal{E}}$ where $\mathcal{E}$ is the process subject to characterization.

\section{Acknowledgements}
The  authors  would  like  to  thank  Dr.   C.  Lombard  Latune  for  insightful  comments  and suggestions. The work  of V.J.  and F.P.  is based upon  research supported  by the
South African Research  Chair Initiative of the  Department of Science
and  Technology and  National Research  Foundation.  R.S. acknowledges the support from Interdisciplinary Cyber Physical Systems (ICPS) Programme of the Department of Science and Technology (DST), India, Grant No: DST/ICPS/QuST/Theme-1/2019/6.

\bibliography{tomoquant}

\begin{thebibliography}{34}%
\makeatletter
\providecommand \@ifxundefined [1]{%
 \@ifx{#1\undefined}
}%
\providecommand \@ifnum [1]{%
 \ifnum #1\expandafter \@firstoftwo
 \else \expandafter \@secondoftwo
 \fi
}%
\providecommand \@ifx [1]{%
 \ifx #1\expandafter \@firstoftwo
 \else \expandafter \@secondoftwo
 \fi
}%
\providecommand \natexlab [1]{#1}%
\providecommand \enquote  [1]{``#1''}%
\providecommand \bibnamefont  [1]{#1}%
\providecommand \bibfnamefont [1]{#1}%
\providecommand \citenamefont [1]{#1}%
\providecommand \href@noop [0]{\@secondoftwo}%
\providecommand \href [0]{\begingroup \@sanitize@url \@href}%
\providecommand \@href[1]{\@@startlink{#1}\@@href}%
\providecommand \@@href[1]{\endgroup#1\@@endlink}%
\providecommand \@sanitize@url [0]{\catcode `\\12\catcode `\$12\catcode
  `\&12\catcode `\#12\catcode `\^12\catcode `\_12\catcode `\%12\relax}%
\providecommand \@@startlink[1]{}%
\providecommand \@@endlink[0]{}%
\providecommand \url  [0]{\begingroup\@sanitize@url \@url }%
\providecommand \@url [1]{\endgroup\@href {#1}{\urlprefix }}%
\providecommand \urlprefix  [0]{URL }%
\providecommand \Eprint [0]{\href }%
\providecommand \doibase [0]{http://dx.doi.org/}%
\providecommand \selectlanguage [0]{\@gobble}%
\providecommand \bibinfo  [0]{\@secondoftwo}%
\providecommand \bibfield  [0]{\@secondoftwo}%
\providecommand \translation [1]{[#1]}%
\providecommand \BibitemOpen [0]{}%
\providecommand \bibitemStop [0]{}%
\providecommand \bibitemNoStop [0]{.\EOS\space}%
\providecommand \EOS [0]{\spacefactor3000\relax}%
\providecommand \BibitemShut  [1]{\csname bibitem#1\endcsname}%
\let\auto@bib@innerbib\@empty
\bibitem [{\citenamefont {Paris}\ and\ \citenamefont
  {Rehacek}(2004)}]{paris_quantum_2004}%
  \BibitemOpen
  \bibinfo {editor} {\bibfnamefont {M.}~\bibnamefont {Paris}}\ and\ \bibinfo
  {editor} {\bibfnamefont {J.}~\bibnamefont {Rehacek}},\ eds.,\ \href@noop {}
  {\emph {\bibinfo {title} {Quantum State Estimation}}},\ \bibinfo {series}
  {Lecture Notes in Physics}\ No.\ \bibinfo {number} {649}\ (\bibinfo
  {publisher} {Springer},\ \bibinfo {address} {Berlin},\ \bibinfo {year}
  {2004})\BibitemShut {NoStop}%
\bibitem [{\citenamefont {Nielsen}\ and\ \citenamefont
  {Chuang}(2000)}]{nielsen_quantum_2000}%
  \BibitemOpen
  \bibfield  {author} {\bibinfo {author} {\bibfnamefont {M.~A.}\ \bibnamefont
  {Nielsen}}\ and\ \bibinfo {author} {\bibfnamefont {I.~L.}\ \bibnamefont
  {Chuang}},\ }\href@noop {} {\emph {\bibinfo {title} {Quantum Computation and
  Quantum Information}}}\ (\bibinfo  {publisher} {Cambridge University Press},\
  \bibinfo {address} {Cambridge},\ \bibinfo {year} {2000})\BibitemShut
  {NoStop}%
\bibitem [{\citenamefont {Jagadish}\ and\ \citenamefont
  {Petruccione}(2018)}]{jagadish_invitation_2018}%
  \BibitemOpen
  \bibfield  {author} {\bibinfo {author} {\bibfnamefont {V.}~\bibnamefont
  {Jagadish}}\ and\ \bibinfo {author} {\bibfnamefont {F.}~\bibnamefont
  {Petruccione}},\ }\href {\doibase 10.12743/quanta.v7i1.77} {\bibfield
  {journal} {\bibinfo  {journal} {Quanta}\ }\textbf {\bibinfo {volume} {7}},\
  \bibinfo {pages} {54} (\bibinfo {year} {2018})}\BibitemShut {NoStop}%
\bibitem [{\citenamefont {Shabani}\ \emph
  {et~al.}(2011{\natexlab{a}})\citenamefont {Shabani}, \citenamefont {Kosut},
  \citenamefont {Mohseni}, \citenamefont {Rabitz}, \citenamefont {Broome},
  \citenamefont {Almeida}, \citenamefont {Fedrizzi},\ and\ \citenamefont
  {White}}]{shabani_efficient_2011}%
  \BibitemOpen
  \bibfield  {author} {\bibinfo {author} {\bibfnamefont {A.}~\bibnamefont
  {Shabani}}, \bibinfo {author} {\bibfnamefont {R.~L.}\ \bibnamefont {Kosut}},
  \bibinfo {author} {\bibfnamefont {M.}~\bibnamefont {Mohseni}}, \bibinfo
  {author} {\bibfnamefont {H.}~\bibnamefont {Rabitz}}, \bibinfo {author}
  {\bibfnamefont {M.~A.}\ \bibnamefont {Broome}}, \bibinfo {author}
  {\bibfnamefont {M.~P.}\ \bibnamefont {Almeida}}, \bibinfo {author}
  {\bibfnamefont {A.}~\bibnamefont {Fedrizzi}}, \ and\ \bibinfo {author}
  {\bibfnamefont {A.~G.}\ \bibnamefont {White}},\ }\href {\doibase
  10.1103/PhysRevLett.106.100401} {\bibfield  {journal} {\bibinfo  {journal}
  {Phys. Rev. Lett.}\ }\textbf {\bibinfo {volume} {106}},\ \bibinfo {pages}
  {100401} (\bibinfo {year} {2011}{\natexlab{a}})}\BibitemShut {NoStop}%
\bibitem [{\citenamefont {Cand{\`e}s}\ \emph {et~al.}(2006)\citenamefont
  {Cand{\`e}s}, \citenamefont {Romberg},\ and\ \citenamefont
  {Tao}}]{candes_stable_2006}%
  \BibitemOpen
  \bibfield  {author} {\bibinfo {author} {\bibfnamefont {E.~J.}\ \bibnamefont
  {Cand{\`e}s}}, \bibinfo {author} {\bibfnamefont {J.~K.}\ \bibnamefont
  {Romberg}}, \ and\ \bibinfo {author} {\bibfnamefont {T.}~\bibnamefont
  {Tao}},\ }\href {\doibase 10.1002/cpa.20124} {\bibfield  {journal} {\bibinfo
  {journal} {Comm. Pure Appl. Math.}\ }\textbf {\bibinfo {volume} {59}},\
  \bibinfo {pages} {1207} (\bibinfo {year} {2006})}\BibitemShut {NoStop}%
\bibitem [{\citenamefont {Donoho}\ and\ \citenamefont
  {Elad}(2003)}]{donoho_optimally_2003}%
  \BibitemOpen
  \bibfield  {author} {\bibinfo {author} {\bibfnamefont {D.~L.}\ \bibnamefont
  {Donoho}}\ and\ \bibinfo {author} {\bibfnamefont {M.}~\bibnamefont {Elad}},\
  }\href {\doibase 10.1073/pnas.0437847100} {\bibfield  {journal} {\bibinfo
  {journal} {Proc. Natl. Acad. Sci. USA}\ }\textbf {\bibinfo {volume} {100}},\
  \bibinfo {pages} {2197} (\bibinfo {year} {2003})}\BibitemShut {NoStop}%
\bibitem [{\citenamefont {Baraniuk}\ \emph {et~al.}(2008)\citenamefont
  {Baraniuk}, \citenamefont {Davenport}, \citenamefont {DeVore},\ and\
  \citenamefont {Wakin}}]{baraniuk_simple_2008}%
  \BibitemOpen
  \bibfield  {author} {\bibinfo {author} {\bibfnamefont {R.}~\bibnamefont
  {Baraniuk}}, \bibinfo {author} {\bibfnamefont {M.}~\bibnamefont {Davenport}},
  \bibinfo {author} {\bibfnamefont {R.}~\bibnamefont {DeVore}}, \ and\ \bibinfo
  {author} {\bibfnamefont {M.}~\bibnamefont {Wakin}},\ }\href {\doibase
  10.1007/s00365-007-9003-x} {\bibfield  {journal} {\bibinfo  {journal}
  {Constr. Approx.}\ }\textbf {\bibinfo {volume} {28}},\ \bibinfo {pages} {253}
  (\bibinfo {year} {2008})}\BibitemShut {NoStop}%
\bibitem [{\citenamefont {Rodionov}\ \emph {et~al.}(2014)\citenamefont
  {Rodionov}, \citenamefont {Veitia}, \citenamefont {Barends}, \citenamefont
  {Kelly}, \citenamefont {Sank}, \citenamefont {Wenner}, \citenamefont
  {Martinis}, \citenamefont {Kosut},\ and\ \citenamefont
  {Korotkov}}]{rodionov_compressed_2014}%
  \BibitemOpen
  \bibfield  {author} {\bibinfo {author} {\bibfnamefont {A.~V.}\ \bibnamefont
  {Rodionov}}, \bibinfo {author} {\bibfnamefont {A.}~\bibnamefont {Veitia}},
  \bibinfo {author} {\bibfnamefont {R.}~\bibnamefont {Barends}}, \bibinfo
  {author} {\bibfnamefont {J.}~\bibnamefont {Kelly}}, \bibinfo {author}
  {\bibfnamefont {D.}~\bibnamefont {Sank}}, \bibinfo {author} {\bibfnamefont
  {J.}~\bibnamefont {Wenner}}, \bibinfo {author} {\bibfnamefont {J.~M.}\
  \bibnamefont {Martinis}}, \bibinfo {author} {\bibfnamefont {R.~L.}\
  \bibnamefont {Kosut}}, \ and\ \bibinfo {author} {\bibfnamefont {A.~N.}\
  \bibnamefont {Korotkov}},\ }\href {\doibase 10.1103/PhysRevB.90.144504}
  {\bibfield  {journal} {\bibinfo  {journal} {Phys. Rev. B}\ }\textbf {\bibinfo
  {volume} {90}},\ \bibinfo {pages} {144504} (\bibinfo {year}
  {2014})}\BibitemShut {NoStop}%
\bibitem [{\citenamefont {Shabani}\ \emph
  {et~al.}(2011{\natexlab{b}})\citenamefont {Shabani}, \citenamefont {Mohseni},
  \citenamefont {Lloyd}, \citenamefont {Kosut},\ and\ \citenamefont
  {Rabitz}}]{shabani_estimation_2011}%
  \BibitemOpen
  \bibfield  {author} {\bibinfo {author} {\bibfnamefont {A.}~\bibnamefont
  {Shabani}}, \bibinfo {author} {\bibfnamefont {M.}~\bibnamefont {Mohseni}},
  \bibinfo {author} {\bibfnamefont {S.}~\bibnamefont {Lloyd}}, \bibinfo
  {author} {\bibfnamefont {R.~L.}\ \bibnamefont {Kosut}}, \ and\ \bibinfo
  {author} {\bibfnamefont {H.}~\bibnamefont {Rabitz}},\ }\href {\doibase
  10.1103/PhysRevA.84.012107} {\bibfield  {journal} {\bibinfo  {journal} {Phys.
  Rev. A}\ }\textbf {\bibinfo {volume} {84}},\ \bibinfo {pages} {012107}
  (\bibinfo {year} {2011}{\natexlab{b}})}\BibitemShut {NoStop}%
\bibitem [{\citenamefont {Rudinger}\ and\ \citenamefont
  {Joynt}(2015)}]{rudinger_compressed_2015}%
  \BibitemOpen
  \bibfield  {author} {\bibinfo {author} {\bibfnamefont {K.}~\bibnamefont
  {Rudinger}}\ and\ \bibinfo {author} {\bibfnamefont {R.}~\bibnamefont
  {Joynt}},\ }\href {\doibase 10.1103/PhysRevA.92.052322} {\bibfield  {journal}
  {\bibinfo  {journal} {Phys. Rev. A}\ }\textbf {\bibinfo {volume} {92}},\
  \bibinfo {pages} {052322} (\bibinfo {year} {2015})}\BibitemShut {NoStop}%
\bibitem [{\citenamefont {Cand{\`e}s}\ and\ \citenamefont
  {Recht}(2009)}]{candes_exact_2009}%
  \BibitemOpen
  \bibfield  {author} {\bibinfo {author} {\bibfnamefont {E.~J.}\ \bibnamefont
  {Cand{\`e}s}}\ and\ \bibinfo {author} {\bibfnamefont {B.}~\bibnamefont
  {Recht}},\ }\href {\doibase 10.1007/s10208-009-9045-5} {\bibfield  {journal}
  {\bibinfo  {journal} {Found. Comput. Math.}\ }\textbf {\bibinfo {volume}
  {9}},\ \bibinfo {pages} {717} (\bibinfo {year} {2009})}\BibitemShut {NoStop}%
\bibitem [{\citenamefont {Recht}(2011)}]{recht2011simpler}%
  \BibitemOpen
  \bibfield  {author} {\bibinfo {author} {\bibfnamefont {B.}~\bibnamefont
  {Recht}},\ }\href {http://www.jmlr.org/papers/v12/recht11a.html} {\bibfield
  {journal} {\bibinfo  {journal} {J. Mach. Learn. Res.}\ }\textbf {\bibinfo
  {volume} {12}},\ \bibinfo {pages} {3413} (\bibinfo {year}
  {2011})}\BibitemShut {NoStop}%
\bibitem [{\citenamefont {Recht}\ \emph {et~al.}(2010)\citenamefont {Recht},
  \citenamefont {Fazel},\ and\ \citenamefont
  {Parrilo}}]{recht_guaranteed_2010}%
  \BibitemOpen
  \bibfield  {author} {\bibinfo {author} {\bibfnamefont {B.}~\bibnamefont
  {Recht}}, \bibinfo {author} {\bibfnamefont {M.}~\bibnamefont {Fazel}}, \ and\
  \bibinfo {author} {\bibfnamefont {P.~A.}\ \bibnamefont {Parrilo}},\ }\href
  {\doibase 10.1137/070697835} {\bibfield  {journal} {\bibinfo  {journal} {SIAM
  Rev.}\ }\textbf {\bibinfo {volume} {52}},\ \bibinfo {pages} {471} (\bibinfo
  {year} {2010})}\BibitemShut {NoStop}%
\bibitem [{\citenamefont {Gross}\ \emph {et~al.}(2010)\citenamefont {Gross},
  \citenamefont {Liu}, \citenamefont {Flammia}, \citenamefont {Becker},\ and\
  \citenamefont {Eisert}}]{gross_quantum_2010}%
  \BibitemOpen
  \bibfield  {author} {\bibinfo {author} {\bibfnamefont {D.}~\bibnamefont
  {Gross}}, \bibinfo {author} {\bibfnamefont {Y.-K.}\ \bibnamefont {Liu}},
  \bibinfo {author} {\bibfnamefont {S.~T.}\ \bibnamefont {Flammia}}, \bibinfo
  {author} {\bibfnamefont {S.}~\bibnamefont {Becker}}, \ and\ \bibinfo {author}
  {\bibfnamefont {J.}~\bibnamefont {Eisert}},\ }\href {\doibase
  10.1103/PhysRevLett.105.150401} {\bibfield  {journal} {\bibinfo  {journal}
  {Phys. Rev. Lett.}\ }\textbf {\bibinfo {volume} {105}},\ \bibinfo {pages}
  {150401} (\bibinfo {year} {2010})}\BibitemShut {NoStop}%
\bibitem [{\citenamefont {Flammia}\ \emph {et~al.}(2012)\citenamefont
  {Flammia}, \citenamefont {Gross}, \citenamefont {Liu},\ and\ \citenamefont
  {Eisert}}]{flammia_quantum_2012}%
  \BibitemOpen
  \bibfield  {author} {\bibinfo {author} {\bibfnamefont {S.~T.}\ \bibnamefont
  {Flammia}}, \bibinfo {author} {\bibfnamefont {D.}~\bibnamefont {Gross}},
  \bibinfo {author} {\bibfnamefont {Y.-K.}\ \bibnamefont {Liu}}, \ and\
  \bibinfo {author} {\bibfnamefont {J.}~\bibnamefont {Eisert}},\ }\href
  {\doibase 10.1088/1367-2630/14/9/095022} {\bibfield  {journal} {\bibinfo
  {journal} {New J. Phys.}\ }\textbf {\bibinfo {volume} {14}},\ \bibinfo
  {pages} {095022} (\bibinfo {year} {2012})}\BibitemShut {NoStop}%
\bibitem [{\citenamefont {Riofr{\'i}­o}\ \emph {et~al.}(2017)\citenamefont
  {Riofr{\'i}­o}, \citenamefont {Gross}, \citenamefont {Flammia}, \citenamefont
  {Monz}, \citenamefont {Nigg}, \citenamefont {Blatt},\ and\ \citenamefont
  {Eisert}}]{riofrio_experimental_2017}%
  \BibitemOpen
  \bibfield  {author} {\bibinfo {author} {\bibfnamefont {C.~A.}\ \bibnamefont
  {Riofr{\'i}­o}}, \bibinfo {author} {\bibfnamefont {D.}~\bibnamefont {Gross}},
  \bibinfo {author} {\bibfnamefont {S.~T.}\ \bibnamefont {Flammia}}, \bibinfo
  {author} {\bibfnamefont {T.}~\bibnamefont {Monz}}, \bibinfo {author}
  {\bibfnamefont {D.}~\bibnamefont {Nigg}}, \bibinfo {author} {\bibfnamefont
  {R.}~\bibnamefont {Blatt}}, \ and\ \bibinfo {author} {\bibfnamefont
  {J.}~\bibnamefont {Eisert}},\ }\href {\doibase 10.1038/ncomms15305}
  {\bibfield  {journal} {\bibinfo  {journal} {Nat. Commun.}\ }\textbf {\bibinfo
  {volume} {8}},\ \bibinfo {pages} {15305} (\bibinfo {year}
  {2017})}\BibitemShut {NoStop}%
\bibitem [{\citenamefont {Steffens}\ \emph {et~al.}(2017)\citenamefont
  {Steffens}, \citenamefont {Riofr{\'i}­o}, \citenamefont {McCutcheon},
  \citenamefont {Roth}, \citenamefont {Bell}, \citenamefont {McMillan},
  \citenamefont {Tame}, \citenamefont {Rarity},\ and\ \citenamefont
  {Eisert}}]{steffens_experimentally_2017}%
  \BibitemOpen
  \bibfield  {author} {\bibinfo {author} {\bibfnamefont {A.}~\bibnamefont
  {Steffens}}, \bibinfo {author} {\bibfnamefont {C.~A.}\ \bibnamefont
  {Riofr{\'i}­o}}, \bibinfo {author} {\bibfnamefont {W.}~\bibnamefont
  {McCutcheon}}, \bibinfo {author} {\bibfnamefont {I.}~\bibnamefont {Roth}},
  \bibinfo {author} {\bibfnamefont {B.~A.}\ \bibnamefont {Bell}}, \bibinfo
  {author} {\bibfnamefont {A.}~\bibnamefont {McMillan}}, \bibinfo {author}
  {\bibfnamefont {M.~S.}\ \bibnamefont {Tame}}, \bibinfo {author}
  {\bibfnamefont {J.~G.}\ \bibnamefont {Rarity}}, \ and\ \bibinfo {author}
  {\bibfnamefont {J.}~\bibnamefont {Eisert}},\ }\href {\doibase
  10.1088/2058-9565/aa6ae2} {\bibfield  {journal} {\bibinfo  {journal} {Quantum
  Sci. Technol.}\ }\textbf {\bibinfo {volume} {2}},\ \bibinfo {pages} {025005}
  (\bibinfo {year} {2017})}\BibitemShut {NoStop}%
\bibitem [{\citenamefont {Liu}\ \emph {et~al.}(2012)\citenamefont {Liu},
  \citenamefont {Zhang}, \citenamefont {Liu}, \citenamefont {Chen},\ and\
  \citenamefont {Yuan}}]{liu_experimental_2012}%
  \BibitemOpen
  \bibfield  {author} {\bibinfo {author} {\bibfnamefont {W.-T.}\ \bibnamefont
  {Liu}}, \bibinfo {author} {\bibfnamefont {T.}~\bibnamefont {Zhang}}, \bibinfo
  {author} {\bibfnamefont {J.-Y.}\ \bibnamefont {Liu}}, \bibinfo {author}
  {\bibfnamefont {P.-X.}\ \bibnamefont {Chen}}, \ and\ \bibinfo {author}
  {\bibfnamefont {J.-M.}\ \bibnamefont {Yuan}},\ }\href {\doibase
  10.1103/PhysRevLett.108.170403} {\bibfield  {journal} {\bibinfo  {journal}
  {Phys. Rev. Lett.}\ }\textbf {\bibinfo {volume} {108}},\ \bibinfo {pages}
  {170403} (\bibinfo {year} {2012})}\BibitemShut {NoStop}%
\bibitem [{\citenamefont {Liu}(2011)}]{liu_universal_2011}%
  \BibitemOpen
  \bibfield  {author} {\bibinfo {author} {\bibfnamefont {Y.-k.}\ \bibnamefont
  {Liu}},\ }in\ \href
  {http://papers.nips.cc/paper/4222-universal-low-rank-matrix-recovery-from-pauli-measurements.pdf}
  {\emph {\bibinfo {booktitle} {Advances in {Neural} {Information} {Processing}
  {Systems} 24}}},\ \bibinfo {editor} {edited by\ \bibinfo {editor}
  {\bibfnamefont {J.}~\bibnamefont {Shawe-Taylor}}, \bibinfo {editor}
  {\bibfnamefont {R.~S.}\ \bibnamefont {Zemel}}, \bibinfo {editor}
  {\bibfnamefont {P.~L.}\ \bibnamefont {Bartlett}}, \bibinfo {editor}
  {\bibfnamefont {F.}~\bibnamefont {Pereira}}, \ and\ \bibinfo {editor}
  {\bibfnamefont {K.~Q.}\ \bibnamefont {Weinberger}}}\ (\bibinfo  {publisher}
  {Curran Associates, Inc.},\ \bibinfo {year} {2011})\ pp.\ \bibinfo {pages}
  {1638--1646}\BibitemShut {NoStop}%
\bibitem [{\citenamefont {Smith}\ \emph {et~al.}(2013)\citenamefont {Smith},
  \citenamefont {Riofr{\'i}­o}, \citenamefont {Anderson}, \citenamefont
  {Sosa-Martinez}, \citenamefont {Deutsch},\ and\ \citenamefont
  {Jessen}}]{smith_quantum_2013}%
  \BibitemOpen
  \bibfield  {author} {\bibinfo {author} {\bibfnamefont {A.}~\bibnamefont
  {Smith}}, \bibinfo {author} {\bibfnamefont {C.~A.}\ \bibnamefont
  {Riofr{\'i}­o}}, \bibinfo {author} {\bibfnamefont {B.~E.}\ \bibnamefont
  {Anderson}}, \bibinfo {author} {\bibfnamefont {H.}~\bibnamefont
  {Sosa-Martinez}}, \bibinfo {author} {\bibfnamefont {I.~H.}\ \bibnamefont
  {Deutsch}}, \ and\ \bibinfo {author} {\bibfnamefont {P.~S.}\ \bibnamefont
  {Jessen}},\ }\href {\doibase 10.1103/PhysRevA.87.030102} {\bibfield
  {journal} {\bibinfo  {journal} {Phys. Rev. A}\ }\textbf {\bibinfo {volume}
  {87}},\ \bibinfo {pages} {030102} (\bibinfo {year} {2013})}\BibitemShut
  {NoStop}%
\bibitem [{\citenamefont {Kyrillidis}\ \emph {et~al.}(2018)\citenamefont
  {Kyrillidis}, \citenamefont {Kalev}, \citenamefont {Park}, \citenamefont
  {Bhojanapalli}, \citenamefont {Caramanis},\ and\ \citenamefont
  {Sanghavi}}]{kyrillidis_provable_2018}%
  \BibitemOpen
  \bibfield  {author} {\bibinfo {author} {\bibfnamefont {A.}~\bibnamefont
  {Kyrillidis}}, \bibinfo {author} {\bibfnamefont {A.}~\bibnamefont {Kalev}},
  \bibinfo {author} {\bibfnamefont {D.}~\bibnamefont {Park}}, \bibinfo {author}
  {\bibfnamefont {S.}~\bibnamefont {Bhojanapalli}}, \bibinfo {author}
  {\bibfnamefont {C.}~\bibnamefont {Caramanis}}, \ and\ \bibinfo {author}
  {\bibfnamefont {S.}~\bibnamefont {Sanghavi}},\ }\href {\doibase
  10.1038/s41534-018-0080-4} {\bibfield  {journal} {\bibinfo  {journal} {npj
  Quantum Inf.}\ }\textbf {\bibinfo {volume} {4}},\ \bibinfo {pages} {36}
  (\bibinfo {year} {2018})}\BibitemShut {NoStop}%
\bibitem [{\citenamefont {Heinosaari}\ \emph {et~al.}(2013)\citenamefont
  {Heinosaari}, \citenamefont {Mazzarella},\ and\ \citenamefont
  {Wolf}}]{heinosaari_quantum_2013}%
  \BibitemOpen
  \bibfield  {author} {\bibinfo {author} {\bibfnamefont {T.}~\bibnamefont
  {Heinosaari}}, \bibinfo {author} {\bibfnamefont {L.}~\bibnamefont
  {Mazzarella}}, \ and\ \bibinfo {author} {\bibfnamefont {M.~M.}\ \bibnamefont
  {Wolf}},\ }\href {\doibase 10.1007/s00220-013-1671-8} {\bibfield  {journal}
  {\bibinfo  {journal} {Commun. Math. Phys.}\ }\textbf {\bibinfo {volume}
  {318}},\ \bibinfo {pages} {355} (\bibinfo {year} {2013})}\BibitemShut
  {NoStop}%
\bibitem [{\citenamefont {Gross}(2011)}]{gross_recovering_2011}%
  \BibitemOpen
  \bibfield  {author} {\bibinfo {author} {\bibfnamefont {D.}~\bibnamefont
  {Gross}},\ }\href {\doibase 10.1109/TIT.2011.2104999} {\bibfield  {journal}
  {\bibinfo  {journal} {IEEE Trans. Inform. Theory}\ }\textbf {\bibinfo
  {volume} {57}},\ \bibinfo {pages} {1548} (\bibinfo {year}
  {2011})}\BibitemShut {NoStop}%
\bibitem [{\citenamefont {Baumgratz}\ \emph {et~al.}(2014)\citenamefont
  {Baumgratz}, \citenamefont {Cramer},\ and\ \citenamefont
  {Plenio}}]{baumgratz_quantifying_2014}%
  \BibitemOpen
  \bibfield  {author} {\bibinfo {author} {\bibfnamefont {T.}~\bibnamefont
  {Baumgratz}}, \bibinfo {author} {\bibfnamefont {M.}~\bibnamefont {Cramer}}, \
  and\ \bibinfo {author} {\bibfnamefont {M.}~\bibnamefont {Plenio}},\ }\href
  {\doibase 10.1103/PhysRevLett.113.140401} {\bibfield  {journal} {\bibinfo
  {journal} {Phys. Rev. Lett.}\ }\textbf {\bibinfo {volume} {113}},\ \bibinfo
  {pages} {140401} (\bibinfo {year} {2014})}\BibitemShut {NoStop}%
\bibitem [{\citenamefont {Greiner}\ and\ \citenamefont
  {M{\"u}ller}(1994)}]{greiner_quantum_1994}%
  \BibitemOpen
  \bibfield  {author} {\bibinfo {author} {\bibfnamefont {W.}~\bibnamefont
  {Greiner}}\ and\ \bibinfo {author} {\bibfnamefont {B.}~\bibnamefont
  {M{\"u}ller}},\ }\href@noop {} {\emph {\bibinfo {title} {Quantum Mechanics -
  Symmetries: With 128 Worked Examples and Problems}}}\ (\bibinfo  {publisher}
  {Springer, Berlin},\ \bibinfo {year} {1994})\BibitemShut {NoStop}%
\bibitem [{\citenamefont {Bechmann-Pasquinucci}\ and\ \citenamefont
  {Tittel}(2000)}]{bechmann-pasquinucci_quantum_2000}%
  \BibitemOpen
  \bibfield  {author} {\bibinfo {author} {\bibfnamefont {H.}~\bibnamefont
  {Bechmann-Pasquinucci}}\ and\ \bibinfo {author} {\bibfnamefont
  {W.}~\bibnamefont {Tittel}},\ }\href {\doibase 10.1103/PhysRevA.61.062308}
  {\bibfield  {journal} {\bibinfo  {journal} {Phys. Rev. A}\ }\textbf {\bibinfo
  {volume} {61}},\ \bibinfo {pages} {062308} (\bibinfo {year}
  {2000})}\BibitemShut {NoStop}%
\bibitem [{\citenamefont {Macchiavello}\ and\ \citenamefont
  {Bruss}(2003)}]{macchiavello_security_2003}%
  \BibitemOpen
  \bibfield  {author} {\bibinfo {author} {\bibfnamefont {C.}~\bibnamefont
  {Macchiavello}}\ and\ \bibinfo {author} {\bibfnamefont {D.}~\bibnamefont
  {Bruss}},\ }\href {\doibase 10.1080/09500340308234549} {\bibfield  {journal}
  {\bibinfo  {journal} {J. Mod. Opt.}\ }\textbf {\bibinfo {volume} {50}},\
  \bibinfo {pages} {1025} (\bibinfo {year} {2003})}\BibitemShut {NoStop}%
\bibitem [{\citenamefont {Durt}\ \emph {et~al.}(2004)\citenamefont {Durt},
  \citenamefont {Kaszlikowski}, \citenamefont {Chen},\ and\ \citenamefont
  {Kwek}}]{durt_security_2004}%
  \BibitemOpen
  \bibfield  {author} {\bibinfo {author} {\bibfnamefont {T.}~\bibnamefont
  {Durt}}, \bibinfo {author} {\bibfnamefont {D.}~\bibnamefont {Kaszlikowski}},
  \bibinfo {author} {\bibfnamefont {J.-L.}\ \bibnamefont {Chen}}, \ and\
  \bibinfo {author} {\bibfnamefont {L.~C.}\ \bibnamefont {Kwek}},\ }\href
  {\doibase 10.1103/PhysRevA.69.032313} {\bibfield  {journal} {\bibinfo
  {journal} {Phys. Rev. A}\ }\textbf {\bibinfo {volume} {69}},\ \bibinfo
  {pages} {032313} (\bibinfo {year} {2004})}\BibitemShut {NoStop}%
\bibitem [{\citenamefont {Grant}\ and\ \citenamefont {Boyd}(2014)}]{cvx}%
  \BibitemOpen
  \bibfield  {author} {\bibinfo {author} {\bibfnamefont {M.}~\bibnamefont
  {Grant}}\ and\ \bibinfo {author} {\bibfnamefont {S.}~\bibnamefont {Boyd}},\
  }\href@noop {} {\enquote {\bibinfo {title} {{CVX}: Matlab software for
  disciplined convex programming, version 2.1},}\ }\bibinfo {howpublished}
  {\url{http://cvxr.com/cvx}} (\bibinfo {year} {2014})\BibitemShut {NoStop}%
\bibitem [{\citenamefont {Jordan}\ and\ \citenamefont
  {Wocjan}(2009)}]{jordan_efficient_2009}%
  \BibitemOpen
  \bibfield  {author} {\bibinfo {author} {\bibfnamefont {S.~P.}\ \bibnamefont
  {Jordan}}\ and\ \bibinfo {author} {\bibfnamefont {P.}~\bibnamefont
  {Wocjan}},\ }\href {\doibase 10.1103/PhysRevA.80.062301} {\bibfield
  {journal} {\bibinfo  {journal} {Phys. Rev. A}\ }\textbf {\bibinfo {volume}
  {80}},\ \bibinfo {pages} {062301} (\bibinfo {year} {2009})}\BibitemShut
  {NoStop}%
\bibitem [{\citenamefont {Childs}(2004)}]{childs2004quantum}%
  \BibitemOpen
  \bibfield  {author} {\bibinfo {author} {\bibfnamefont {A.~M.}\ \bibnamefont
  {Childs}},\ }\emph {\bibinfo {title} {Quantum Information Processing in
  Continuous Time}},\ \href@noop {} {Ph.D. thesis},\ \bibinfo  {school} {MIT}
  (\bibinfo {year} {2004})\BibitemShut {NoStop}%
\bibitem [{\citenamefont {Aharonov}\ and\ \citenamefont
  {Ta-Shma}(2003)}]{aharonov_adiabatic_2003}%
  \BibitemOpen
  \bibfield  {author} {\bibinfo {author} {\bibfnamefont {D.}~\bibnamefont
  {Aharonov}}\ and\ \bibinfo {author} {\bibfnamefont {A.}~\bibnamefont
  {Ta-Shma}},\ }in\ \href {\doibase 10.1145/780542.780546} {\emph {\bibinfo
  {booktitle} {Proceedings of the Thirty-Fifth {ACM} Symposium on {Theory} of
  Computing - {STOC} '03}}}\ (\bibinfo  {publisher} {ACM Press},\ \bibinfo
  {address} {San Diego, CA, USA},\ \bibinfo {year} {2003})\ p.~\bibinfo {pages}
  {20}\BibitemShut {NoStop}%
\bibitem [{\citenamefont
  {Jamio{\l}'kowski}(1974)}]{jamiolkowski_effective_1974}%
  \BibitemOpen
  \bibfield  {author} {\bibinfo {author} {\bibfnamefont {A.}~\bibnamefont
  {Jamio{\l}'kowski}},\ }\href {\doibase
  https://doi.org/10.1016/0034-4877(74)90044-5} {\bibfield  {journal} {\bibinfo
   {journal} {Rep. Math. Phys.}\ }\textbf {\bibinfo {volume} {5}},\ \bibinfo
  {pages} {415 } (\bibinfo {year} {1974})}\BibitemShut {NoStop}%
\bibitem [{\citenamefont {Choi}(1975)}]{choi_completely_1975}%
  \BibitemOpen
  \bibfield  {author} {\bibinfo {author} {\bibfnamefont {M.-D.}\ \bibnamefont
  {Choi}},\ }\href {\doibase https://doi.org/10.1016/0024-3795(75)90075-0}
  {\bibfield  {journal} {\bibinfo  {journal} {Linear Algebr. Appl.}\ }\textbf
  {\bibinfo {volume} {10}},\ \bibinfo {pages} {285 } (\bibinfo {year}
  {1975})}\BibitemShut {NoStop}%
\end{thebibliography}%

\end{document}